\documentclass[nofootinbib,preprint,
showpacs,preprintnumbers,amsmath,amssymb]{revtex4}
\usepackage{latexsym}
\usepackage{hyperref}
\usepackage{epsfig, psfrag,graphicx}
\usepackage{amssymb}
\usepackage{amsmath}

\newcommand{\be}{\begin{equation}}
\newcommand{\ee}{\end{equation}}
\newcommand{\bn}{\begin{equation*}}
\newcommand{\en}{\end{equation*}}
\newcommand{\ben}{\begin{eqnarray*}}
\newcommand{\een}{\end{eqnarray*}}
\newcommand{\bea}{\begin{eqnarray}}
\newcommand{\eea}{\end{eqnarray}}
\newcommand{\bdm}{\begin{displaymath}}
\newcommand{\edm}{\end{displaymath}}
\newcommand{\ba}{\begin{align}}
\newcommand{\ea}{\end{align}}
\newcommand{\lb}{\label}
\newcommand{\D}{{\rm d}}

\begin{document}

\title{\bf MEASUREMENT ANALYSIS AND QUANTUM GRAVITY}

\author{Mark Albers}
\affiliation{Institut f\"ur Theoretische Physik,
Universit\"{a}t zu K\"{o}ln, Z\"{u}lpicher
Strasse 77,
50937 K\"{o}ln, Germany}
\author{Claus Kiefer}
\affiliation{Institut f\"ur Theoretische Physik,
Universit\"{a}t zu K\"{o}ln, Z\"{u}lpicher
Strasse 77,
50937 K\"{o}ln, Germany}
\author{Marcel Reginatto}
\affiliation{Physikalisch-Technische Bundesanstalt,
Bundesallee 100, 38116 Braunschweig, Germany}

\date{\today}

\begin{abstract}

We consider the question of whether consistency arguments based
on measurement theory show that the gravitational field must be
quantized. Motivated by the argument of Eppley and Hannah, we
apply a DeWitt-type measurement analysis to a coupled system that
consists of a gravitational wave interacting with a mass cube.
We also review the arguments of Eppley and Hannah and of DeWitt, and
investigate a second model in which a gravitational wave
interacts with a quantized scalar field. We argue that one cannot
conclude from the existing gedanken experiments that gravity has
to be quantized. Despite the many physical arguments which speak in
favor of a quantum theory of gravity, it appears that the justification
for such a theory must be based on empirical tests and does not follow
from logical arguments alone.

\end{abstract}
\maketitle


\section{Introduction\label{intro}}

The theoretical analysis of the measurement process has played a crucial
role in the development of quantum theory. A famous example is the
discussion between Einstein and Bohr on the role of the uncertainty
relations, which took place during the Solvay conferences in Brussels
at the end of the 1920s. Another example is the analysis of quantum
electrodynamical field quantities by Bohr and Rosenfeld in 1933
\cite{BR}. We are interested in the question of whether an analysis
along these lines could also be of help in the search for a quantum
theory of gravity.

Quantum gravity does not yet exist in a final form, but many
promising, although competing approaches, exist \cite{OUP}. Among them
are string theory, quantum geometrodynamics, loop quantum gravity,
path-integral quantization, and others. It would be of interest to see
which variables can be accessible in a quantum
measurement and whether they can be measured with arbitrary accuracy
or not. In the canonical approaches to quantum gravity, candidates for
such variables are the three-dimensional metric and the second
fundamental form, or holonomies and fluxes of the densitized triad; in
string theory, the candidates for the most fundamental variables are
less clear.

The present paper can be seen as preparation for such an analysis of
quantum gravitational variables. It addresses the question of whether a
measurement analysis could disclose that the gravitational field {\em
  must} be quantized for consistency. There exist various papers in
the literature that claim that this question must be answered in the
affirmative, see for example \cite{DeWitt62} and \cite{EH}.
In our paper, we show that such a conclusion cannot be
drawn. Many physical arguments speak in favour of quantum gravity
(such as the existence of the singularity theorems in the classical theory)
\cite{OUP}, but the final justification can only come from an empirical test
and not from logical arguments alone.

Our paper is organized as follows. In Sec. II, we address the gedanken
experiment discussed in \cite{EH}. We disclose gaps in the chain
of argument which invalidate the conclusion drawn that gravity must be
quantized. In Sec. III, we apply the formalism of quantum
measurement analysis developed by Bryce DeWitt to such a gedanken
experiment. This serves the purpose of putting the heuristic discussion of
\cite{EH} on a quantitative level, but
is also of interest in its own right -- as a study of the relationship
between classical and quantum theory. We show that (and how) the uncertainties
present in one system entail uncertainties in the system to which it
is coupled, but that this does not enforce the quantization of the
coupled system. In Sec. IV, we present an explicit counterexample to the
claim that a system coupled to a quantum system must necessarily also
be of quantum nature: we discuss a hybrid model with a consistent coupling
between classical gravity and a quantized scalar field. In Sec. V, we consider
the argument made in \cite{DeWitt62} that the quantum theory must be extended
to all physical systems and show that this conclusion is not justified.
Sec. VI gives a brief summary and an outlook.


\section{Critique of the Eppley and Hannah gedanken experiment}
\lb{Critique of EH}

In 1977, Eppley and Hannah proposed a gedanken experiment which
was meant to demonstrate that the gravitational field
must be quantized \cite{EH}.
They considered the interaction of a {\em classical} gravitational wave
with a {\em quantum} system, and argued that this would lead to
a violation of momentum conservation, to a violation of the uncertainty
principle, or to the transmission of signals faster than light. Since
nothing special about gravity seems to enter their
line of thought (one is, in particular, at the level of a linearized wave),
their arguments should hold for any classical wave, in particular for an
electromagnetic wave. One thus seems to be led to the conclusion that
any system that is coupled to a quantum system must also be a quantum
system.

A gedanken experiment should, of course, be realizable at least in
principle. It was recently emphasized by Mattingly that this is not the
case for the Eppley--Hannah (EH) model; for example,
their detector must be so massive as to be within its own
Schwarzschild radius \cite{Mattingly}.
Furthermore, it was argued by Huggett and Callender that the
violations of physical principles are only present in the Copenhagen
interpretation of quantum mechanics and are thus, at least partially,
resolvable within alternative interpretations \cite{HC}.
Here we show that even without the question of
realizability or interpretation, it is not correct to say that the EH gedanken
experiment entails the necessity of quantizing the gravitational
field.

Eppley and Hannah consider the interaction of a classical
gravitational wave of small momentum with a quantum particle
described by a wave function $\psi$. They restrict to the case in which
the wavelength, $\lambda$, of the wave is much smaller than the position
uncertainty, $\Delta x$, of the particle in order for the interaction to
lead to a measurement of the position of the quantum particle.
For the generation and detection
of the gravitational wave, they make the following
assumption: the uncertainties of the
quantum system which is used to prepare and detect the wave will
result in a negligible perturbation of the classical wave. To discuss
this assumption, we shall present a DeWitt-type measurement analysis
in the following section. Here we take their assumption for
granted.
When the quantum particle is probed by the classical wave, it is
furthermore assumed that no direct perturbation of the particle occurs
because of the small wave momentum. This constitutes an ``ideal
measurement'' in the sense of John von Neumann,
a condition that is now also called ``quantum non-demolition
measurement'', cf. \cite{deco}.
(We postpone a discussion of this assumption until the end of this
section; here, we also take it for granted.)

Eppley and Hannah now distinguish between two possibilities: either
the gravitational wave leads to a collapse of the wave function of the
particle, or it does not. If it does, they argue that this would
entail either momentum non-conservation or a violation of the
uncertainty relation. If it does not collapse, they argue that there
will be a transmission of information with superluminal speed.

Let us consider the first case of an assumed collapse. Such a collapse
is not part of the standard linear quantum theory because it would
violate the superposition principle. One can, of course, modify
quantum mechanics in order to accommodate such a collapse, cf. Chapter~8
of \cite{deco} or \cite{Pearle}. In fact, one popular class of such
models are models of gravity-induced collapse of the wave function.
Such collapse models typically introduce new
constants of nature. They often have problems with conservation laws,
so some effort is required in order to construct a model that is in
accordance with such laws. It is thus not surprising that EH arrive at
problems with momentum non-conservation, but this by itself should not
be taken as a logical argument to quantize gravity.
Therefore, before any definite statement about a
possible violation of momentum conservation or the uncertainty relations
can be made, the interaction of the gravitational wave with the
quantum particle must be discussed quantitatively within a definite
collapse model. The arguments of \cite{EH} are therefore
inconclusive for this case.

Let us assume, then, that the gravitational wave does not
collapse the wave function $\psi$ of the particle. Eppley and Hannah
then consider an EPR-type situation where one particle decays into two
other particles which together are in a singlett state (e.g. a $\pi^0$
decaying into two $\gamma$): if an observer measures photon 1 in
horizontal polarization, photon 2 will be found in vertical
polarization, and vice versa. Eppley and Hannah then argue that a classical
gravitational wave scattered off from particle 2 can distinguish
between particle 2 having a definite polarization or particle 2 being
in a superposition of both polarizations. The gravitational wave could
thus instantaneously `see' whether a measurement at particle 1 was
done -- the corresponding information would then have propagated with
superluminal speed. This conclusion is, however, not correct. Before
the measurement of particle 1, the total state of photons and detector
is in the entangled state
\be
\lb{singlett}
\vert\Psi_0\rangle=\frac{1}{\sqrt{2}}
\left(\vert\uparrow\rangle_1\vert\downarrow\rangle_2-
\vert\downarrow\rangle_1\vert\uparrow\rangle_2\right)\vert\Phi_0\rangle\ ,
\ee
where $\vert\uparrow\rangle_1$ ($\vert\downarrow\rangle_1 $) denotes
horizontal (vertical) polarization of photon 1 (and similarly for
photon 2), and $\vert\Phi_0\rangle$ denotes the initial
(switched off) state of the
detector which will measure photon 1. After the detector has measured
photon 1, the initial state $\vert\Psi_0\rangle$ will have evolved into
the new entangled state
\be
\lb{afterdetection}
\vert\Psi\rangle=\frac{1}{\sqrt{2}}\left(\vert\uparrow\rangle_1
\vert\downarrow\rangle_2\vert\Phi_{\uparrow}\rangle-
\vert\downarrow\rangle_1\vert\uparrow\rangle_2
\vert\Phi_{\downarrow}\rangle\right)\ ,
\ee
where $\vert\Phi_{\uparrow}\rangle$ ($\vert\Phi_{\downarrow}\rangle$)
denotes the state of the detector after it has measured the
polarization of photon 1 to be horizontal (vertical). The important
point is that photon 2, by itself, is not in a pure state, neither
before nor after the measurement of photon 1. It finds itself in a
mixed state which is obtained from the total pure state of system plus
detector by tracing out the states of particle 1 and the
detector. This leads for {\em both} (\ref{singlett}) and
(\ref{afterdetection}) to the same density operator for photon 2,
\be
\hat{\rho}=\frac12 \left(\vert\uparrow\rangle_2\langle\uparrow\vert_2+
\vert\downarrow\rangle_2\langle\downarrow\vert_2\right)\ .
\ee
In both cases, photon 2, by itself, is in a mixed state of horizontal and
vertical polarization with equal probability one half.
The gravitational wave `sees' the same mixed state for photon 2,
independent of whether a measurement of photon 1 has been performed or
not; thus, no superluminal communication is possible. The
difference between the total states (\ref{singlett}) and (\ref{afterdetection})
can only be seen after both photons are brought together: in case
(\ref{singlett}) they will interfere, while in case
(\ref{afterdetection}) they will not; in the latter case the
information about the original superposition has been delocalized into
correlations of the detector state with its natural environment, that
is, decoherence has occurred \cite{deco}.

The above argument assumes that the linear structure of quantum theory
remains untouched. This, of course, corresponds to an Everett
interpretation. If, on the other hand, one assumes that a measurement
on photon 1 collapses its wave function into $\vert\uparrow\rangle_1$
or $\vert\downarrow\rangle_1$, a superluminal communication might in
principle be possible, cf. \cite{Carlip}.
This can happen, for example, in the context of semiclassical gravity
(see e.g. \cite{OUP}) where the source of the gravitational field is
taken to be the quantum expectation value of the energy--momentum
tensor. Semiclassical gravity introduces a non-linearity
into quantum theory, which in principle can be
experimentelly tested \cite{Carlip}. The possibility of such an
example does not prove, however, that a mixed classical-quantum
coupling without superluminal communication is impossible. In fact,
Sect.~IV presents a candidate for such a theory. In Appendix~B we
shall discuss a simple gedanken experiment of a classical particle
interacting with a quantum system; we shall show there that the
mixed classical-quantum coupling of Sect.~IV can successfully deal
with such a situation. This is possible because the classical and
quantum sectors still exhibit some entanglement.

One of the assumptions that we have taken for granted here, namely,
that the momentum of the wave is sufficiently small for the particle
not to be perturbed, leads to difficulties if the scattering of the
gravitational wave is going to be used to measure the position of the
particle. Let us assume, as Eppley and Hannah do, that the interaction
is such that incoming plane gravitational waves are scattered by the
quantum system with the result that spherical waves are emitted. To
achieve this, a certain amount of energy needs to be transferred from
the incoming gravitational wave to the quantum system, because the
system will not emit gravitational waves unless its quadrupole or
higher multipole modes are excited. Given that a quantum system
typically has quadrupole modes that are quantized, one would expect
this energy to be of the order of $h \nu$, where $\nu$ is some
characteristic frequency associated with the transition to the
quadrupole mode. But if we assume that the incoming gravitational wave
carries a negligible amount of momentum, then it will not be able to
transfer a sufficient amount of energy to the quantum system to excite
its higher modes and no scattering can take place. And if we allow the
gravitational wave to carry sufficient energy for scattering to occur,
then we cannot rule out a transfer of energy and momentum that may be
sufficient for the quantum system to be left in a state that does not
violate the Heisenberg uncertainty relation, independent of any
particular model that might be used to describe collapse of the wave
function.

A more detailed description of the
interaction term is required before we can make more conclusive
statements regarding the outcome of the EH gedanken
experiment. Presumably, the gravitational wave will interact with the
quantized system through gravitational effects only. If so, then the
form of the interaction will be determined by the way in which gravity
couples to matter. In the next sections, we present examples
where these considerations are taken into account and particular
interaction terms are examined.


\section{DeWitt-type measurement analysis}\label{DeWitt m.a.}

\subsection{General analysis}

Motivated, in particular, by the EH gedanken experiment discussed in
the last section, we present a general measurement analysis of an
interaction between a gravitational wave and a non-gravitational
system. For this purpose, we apply the general formalism introduced by
Bryce DeWitt, cf. \cite{DeWitt62}, \cite{DeWJMP}, \cite{DeW}, and
\cite{festschrift}. We consider this type of analysis for two reasons.
First, DeWitt's approach provides a descriptive and straightforward
method of modeling a measurement (or any other influence of one system
on another). Second, it provides a convenient way of developing a
concrete realization of the EH gedanken experiment,
see Sec.~\ref{mass-cube+gw} below.
Before describing our model, we review the
general framework.

The starting point of the measurement analysis developed by DeWitt is the
observation that a measurement of a physical observable requires an
interaction between the system and the apparatus. A measurement is
then interpreted as recording the resulting change in the state of the
apparatus, that is, the difference between the configurations with and
without the interaction taking place. For example, for a voltage
measurement this would be 2 volt if the initial value increases (due
to the coupling between apparatus and system) from 5 to 7 volt.

There are four basic ingredients that enter into the analysis and
which need to be modeled theoretically: the physical system, the
measuring apparatus, the choice of system observable that is being
measured, and the coupling term that describes the interaction between
the system and the apparatus. At this level of description, the focus
is on possible theoretical limitations rather than on the practical
limitations that are unavoidable when carrying out measurements in a
laboratory. Therefore, other complications that are encountered in
real experiments are not considered here.
Once these four ingredients are defined, one can follow DeWitt's
method and calculate the change in the configuration of the apparatus
induced by the coupling to the system. Two assumptions are needed to
make the equations tractable. First, the coupling is assumed to be
weak. This is an essential assumption that allows a
perturbative approach using first and second order terms
only. Moreover, this assumption corresponds to the idea of a careful
measurement which imparts only a small disturbance and is therefore justified when investigating `...uncontrollable uncertainties which
remain in spite of all precautions.' \cite{DeWJMP}. Second, it is
assumed that the second order terms which involve the disturbance of
the apparatus can be neglected. This approximation is valid because of
the usual $1/m^2$ dependence of the Green functions of
the given systems together with the condition that the apparatus
be much more massive than the system. Such an assumption helps simplify the
equations, but one may also take such terms into consideration if
necessary.

The first step in this perturbative analysis is to solve the equations
for the system and apparatus while neglecting the interaction term,
that is, the equations
\begin{eqnarray*}
S[\phi],_i=0 \ ,\\
\Sigma[\theta],_I=0  \ ,
\end{eqnarray*}
where $S[\phi]$ denotes the action functional of the uncoupled system,
$\Sigma[\theta]$ the action functional of the uncoupled
apparatus, $\phi$ are system variables and $\theta$ are apparatus
variables. The comma denotes functional differentiation and lower-case
(capital) latin letters are used to indicate derivatives with respect
to system (apparatus) variable. We make use of DeWitt's condensed
notation, where a summation over discrete indices indicates an
integration over the argument of the field (e.g., given $A_i(x,t)$ and
$B_i(x,t)$, $A_iB^i$ should be read as
$\sum_i\int {\mathrm d}x{\mathrm d}tA_iB^i$). The solutions of the
uncoupled equations will be called $\phi_0$ and $\theta_0$, respectively.

Once the uncoupled field configurations are calculated, one introduces
the interaction term $\Omega[\phi,\theta]$.
The interaction is assumed to last for a finite time only, which means
that all terms describing the perturbations that arise due to the
interaction (i.e., $\delta\phi$, $\delta\theta$, etc.) must satisfy
{\em retarded} boundary conditions,
\begin{eqnarray}
\lim_{t\rightarrow-\infty}\delta\phi(t)=0 \nonumber\ ,\\
\lim_{t\rightarrow-\infty}\delta\theta(t)=0 \label{ret.condition}\ .
\end{eqnarray}
If we take into consideration the interaction term, the total action functional takes the form
\begin{equation}
S[\phi]+\Sigma[\theta]\rightarrow S[\phi] + \Sigma[\theta]+
 g \Omega[\phi,\theta]\lb{tot.unc.act.} \ ,
\end{equation}
with a small dimensionless coupling constant $g$.

We now introduce the assumption that the coupling is weak and expand
the equations derived from \eqref{tot.unc.act.} around the solutions
$\phi_0$ and $\theta_0$ of
the uncoupled equations,
\begin{eqnarray*}
S,_i[\phi_0+\delta\phi]+
g\Omega,_i[\phi_0+\delta\phi,\theta_0+\delta\theta]=0\ ,\\
\Sigma,_I[\theta_0+\delta\theta]+
g\Omega,_I[\phi_0+\delta\phi,\theta_0+\delta\theta]=0\ ,
\end{eqnarray*}
where due to the small coupling the stationary points are expected to
lie close to $\phi_0$ and $\theta_0$.
This leads to the functional Taylor series
\begin{eqnarray}
& S,_i[\phi_0]+S,_{ij}[\phi_0]\delta\phi^j+\frac{1}{2}S,_{ijk}[\phi_0]\delta\phi^j\delta\phi^k+\nonumber\\
 &
 g\Omega,_i[\phi_0,\theta_0]+g\Omega,_{ij}[\phi_0,\theta_0]\delta\phi^j+g\Omega,_{iI}[\phi_0,\theta_0]\delta\theta^I+\cdots=0 \lb{exp.of pat.}
\end{eqnarray}
and
\begin{eqnarray}
& \Sigma,_I[\theta_0]+\Sigma,_{IJ}[\theta_0]\delta\theta^J+\frac{1}{2}\Sigma,_{IJK}[\theta_0]\delta\theta^J\delta\theta^K+\nonumber\\
 & g\Omega,_I[\phi_0,\theta_0]+g\Omega,_{IJ}[\phi_0,\theta_0]\delta\theta^J+g\Omega,_{Ii}[\phi_0,\theta_0]\delta\phi^i+\cdots=0 \ . \lb{exp.of ap.}
\end{eqnarray}
In principle, equations \eqref{exp.of pat.} and \eqref{exp.of ap.} can
be used to derive solutions to any order. Within DeWitt's
scheme one has to calculate from \eqref{exp.of pat.} the change in the
configuration of the system only to first order, that is, to solve
\begin{equation}
\lb{Sij}
S,_{ij}[\phi_0]\delta\phi^j=-g\Omega,_i[\phi_0,\theta_0] \ .
\end{equation}
Back reaction of the apparatus change onto the system
(which would correspond to the term $g\Omega,_{iI}\delta\theta^I$) is
thus not considered.
Note that terms involving the first functional derivative of the
action vanish because $\phi_0$ and $\theta_0$ are solutions of the
uncoupled equations
of motion. To get an expression for the apparatus which takes into account
the system changes, it is necessary to solve equation
\eqref{exp.of ap.} to second order. The calculation simplifies
if we now make use of the second assumption and neglect all second-order
terms involving $\delta\theta$ and any $\delta\theta$ terms that
appear together with $\delta\phi$ terms of the same order. When this
is done, we are led from \eqref{exp.of ap.} to the equations
\begin{equation}
\lb{SigmaIJ}
\Sigma,_{IJ}[\theta_0]\delta\theta^J=
-g\Omega,_I[\phi_0,\theta_0]-g\Omega,_{Ii}[\phi_0,\theta_0]\delta\phi^i  \ .
\end{equation}
The last term describes the back reaction of the system on the
apparatus. It is also possible to have the special case where the
interaction between system and apparatus causes no system disturbance,
that is, with $\delta\phi^i=0$. In this case, one needs to use
\eqref{exp.of ap.} without the above approximation of neglecting certain
terms in $\delta\theta$.

Finally, one needs to choose an appropriate apparatus observable, $A$,
to calculate the change, $\delta A$, which arises due to the
disturbance, $\delta\theta$, within the apparatus configuration,
\be
\lb{deltaA}
\delta A=A,_I\delta\theta^I \ .
\ee
While the brief description given here does not present DeWitt's
original procedure in full (which makes use of Peierls brackets and
Green's functions, see Sec. \ref{CQ-discussion}), it is sufficient
for the analysis presented in the next subsection, where we apply the
formalism to a concrete model. Note that so far the analysis is
entirely classical.

\subsection{Mass-cube interacting with a gravitational wave packet}\lb{mass-cube+gw}

The motivation for this model arose from the critique of \cite{EH}
presented in Sec. \ref{Critique of EH}. While the actual results
and interpretations at the end of this subsection are heuristic, it will
be shown that they support our critique mentioned before.

We consider a cube with homogeneous mass density interacting with a
linear gravitational wave packet. The cube is characterized by its
edge length, $a$, and its mass, $M$. Its speed is assumed to be small
compared to the speed of light (non-relativistic approximation). It
can be described by the action
\begin{equation}
\lb{systemaction}
S=\int \D t\ \frac12 M\dot{\mathbf r}^2=\frac12
\int \D ^4x\  \rho \dot{\mathbf r}^2\ .
\end{equation}
The mass density of the cube is given by
\begin{eqnarray}
\label{cubedensity}
\rho&=&\frac{M}{a^3}\Theta\left(x_1 -x_{1_{\rm cm}}(t)+\frac a2\right)
\left[ 1 - \Theta\left(x_1-x_{1_{\rm cm}}(t)-\frac a2\right) \right]\times \nonumber \\
 &\,&\;\quad\Theta\left(x_2 -x_{2_{\rm cm}}(t)+\frac a2\right)
   \left[ 1 - \Theta\left(x_2 -x_{2_{\rm cm}}(t)-\frac a2\right)\right]\times \nonumber \\
&\,&\quad\;\Theta\left(x_3 -x_{3_{\rm cm}}(t)+\frac a2\right)
  \left[ 1 - \Theta\left(x_3 -x_{3_{\rm cm}}(t)-\frac a2\right)\right] \ ;
\end{eqnarray}
the subscript `cm' is an abbreviation for `center of mass', and the
$x_{i_{\rm cm}}$ are the time-dependent coordinates of the center of
mass of the cube, that is, they correspond to the aforementioned dynamical
variable of the system $\phi$. A possible solution of the free
equation of motion, that is, the solution corresponding to the $\phi_0$
above, is
\begin{equation}
\mathbf r_{\rm cm}(t)=(0,vt,0)\ ,\label{free system}
\end{equation}
which is unaccelerated motion in $x_2$-direction with velocity
$v$.

If the cube consists of non-interacting particles, the components of
its stress--energy tensor read
\begin{eqnarray*}
T_{00}&=& \rho c^2\nonumber \\
T_{ik} &=& \rho u_{i} u_{k}=\rho \dot{x}_{i_{\rm cm}}(t)\dot{x}_{k_{\rm cm}}(t)\ ,
\end{eqnarray*}
where $\rho$ is given by (\ref{cubedensity}). (We have neglected here
possible internal stresses of the cube, so the cube is in first
approximation interpreted as a set of non-interacting particles.)
The linear gravitational wave packet is, as usual, a superposition of
plane waves, which are solutions of the linear Einstein
equations. In the spirit of the EH-model, the gravitational wave plays
here the role of the `apparatus'.
The Einstein equations follow from the variation of the
Einstein--Hilbert action
\begin{eqnarray*}
\Sigma[g_{\mu \nu}]&=&\int d^4x\frac{\sqrt{-g}R}{2\kappa}\ ,
\end{eqnarray*}
where $\kappa={8\pi G}/{c^4}$. For the limit of linearized gravity one gets
\begin{equation}
\Box\Psi_{\mu \nu}=0\lb{eom.app.} \ .
\end{equation}
Here, $g_{\mu \nu} \approx \eta_{\mu \nu} + 2 h_{\mu \nu}$ with
$|h_{\mu \nu}|\ll 1$, $\Psi_{\mu \nu}\equiv
h_{\mu\nu}-\frac{1}{2}\eta_{\mu \nu} h_{\lambda}^{\lambda}$,
$\eta_{\mu \nu}={\rm diag}\ (-1,1,1,1)$, and we use the harmonic
gauge which corresponds to $\Psi_{\mu \nu},^{\nu}=0$. We are interested
in solutions of \eqref{eom.app.} that represent gravitational waves
propagating into a direction of maximal coupling (perpendicular
to the path of the cube), for example, the $x_1$-direction:
\be
\Psi^{\omega_0}_{\mu \nu}(x)=\mathrm{Re} \left( A_+ e_{\mu
    \nu}^+e^{-i\omega_0\left(t-\frac{x_1}{c}\right)}\right),
\label{free apparatus}
\ee
where $e_{\mu \nu}^+$ denotes a polarization tensor which reads
\begin{eqnarray*}
e_{\mu \nu}^+=\left(\begin{array}{l c c r}
0&0&0&0\\
0&0&0&0\\
0&0&1&0\\
0&0&0&-1
\end{array}\right) \ ;
\end{eqnarray*}
this describes the $+$-polarization, see, for example, \cite{MTW},
Chap.~35. A superposition of such solutions with Gaussians
$A_0\exp(-b(\omega_0-\omega)^2/2)$ according to
\begin{eqnarray}
\Psi_{\mu \nu}(x_1,t)&=&\mathrm{Re}\big{[}e_{\mu\nu}^+ A_+\int d\omega_0 A_0 e^{-\frac{b}{2}(\omega_0-\omega)^2-i\omega_0\left(t-\frac{x_1}{c}\right)}\big{]}\nonumber\\
&\equiv&\mathrm{Re}\big{[}e_{\mu\nu}^+ A\sqrt{\frac{2\pi}{b}}
e^{-\frac{1}{2b}\left(t-\frac{x_1}{c}\right)^2-i\omega\left(t-\frac{x_1}{c}\right)}\big{]},
\label{PSI}
\end{eqnarray}
where $A=A_+A_0$, yields the requested linear gravitational wave packet.
We note that this solution $\Psi_{\mu \nu}$ corresponds to
$\theta_0$ in Sec. III.A.

The coupling between wave and cube is modeled via a linear interaction of
the wave and the stress--energy tensor of the mass contribution. This
constitutes a non-linear coupling between the gravitational field and
the particles, and thus goes beyond the usual limit in which the
particles are only considered as test particles in an external
gravitational wave.
The interaction is switched on at $t=T$ in order to follow
DeWitt's procedure that all disturbances have to obey equation
\eqref{ret.condition}; hence we assume
\be
\lb{interaction}
\Omega[\mathbf r(t),\Psi_{\mu\nu}(x_1,t)]=
\int d^4x T_{\mu\nu}\Psi^{\mu\nu}\theta(t-T)\ .
\ee
The contraction of $T_{\mu\nu}$ with $\Psi^{\mu\nu}$ in this form
of interaction defines which states of motion will yield
nonvanishing contributions. A brief look at $e_{\mu \nu}^+$ shows that
only the $T_{22}$ and $T_{33}$ components will couple to the
scattering gravitational wave. Moreover, they are going to contribute
with opposite sign. This indicates that certain systems, for example
an isotropic scalar field, would
not couple to such a gravitational wave because the
interaction would be proportional to $k^2_2-k^2_3=k^2_2-k^2_2=0$.

For this model, the interaction causes disturbances $\delta
\mathbf{r}_{\rm cm}$ and $\delta\Psi$ of the initially free solutions
\eqref{free system} and \eqref{free apparatus} which, if assumed
small, are given by the general expressions \eqref{Sij} and
\eqref{SigmaIJ}. Following the
procedure described above (i.e., to take the changes within the system
into account in order to calculate those of the apparatus), we first
compute $\delta\mathbf{r}_{\rm cm}$.

We shall use \eqref{Sij} with $x_j$ for the $\phi^j$ and
$\psi_{\mu\nu}$ for the $\theta^J$; we set $g=1$. Then,
\be
S_{,ij}\delta x^j=-M\delta\ddot{x}^i(t),
\ee
and, since $T_{\mu\nu}\psi^{\mu\nu}=\rho\dot{x}_2^2\psi^{22}$,
\be
\lb{Omegaxi}
-\frac{\delta\Omega}{\delta x^i(t)}=-\int
d^3x\left[\frac{\partial\rho}{\partial x^i}\dot{x}_2^2\psi^{22}-
\delta^i_2\frac{d}{dt}\left(2\rho\dot{x}_2\psi^{22}\right)\right]\theta(t-T).
\ee
We shall consider first the disturbance in
$x_2(t)$.
Equation \eqref{Sij} reads for this case
\be
\lb{Sij2}
-M\delta\ddot{x}_2(t)=-\frac{\delta\Omega}{\delta x^2(t)}.
\ee
For the right-hand side we find that only the second term in
\eqref{Omegaxi} contributes. Setting $T=0$, we then get
\bea
\lb{Omegax2}
-\frac{\delta\Omega}{\delta x^2(t)} &=& \int d^3x\left[\frac{d}{dt}
\left(2\rho\dot{x}_2\psi^{22}\right)\right]\theta(t)\nonumber\\
& = & -\frac{2Mvc}{a}\theta(t)\left(\psi_+^{22}-\psi_-^{22}\right)-
\frac{2Mv}{a}\delta(t)\int_{-a/2}^{a/2}dx_1\psi^{22}(x_1,t),
\eea
where
\ben
\psi_+^{22} &=& A\sqrt{\frac{2\pi}{b}}e^{-\frac{t_+^2}{2b}-i\omega
  t_+},\\
\psi_-^{22} &=& A\sqrt{\frac{2\pi}{b}}e^{-\frac{t_-^2}{2b}-i\omega
  t_-},
\een
and $t_+=t+a/2c$ and $t_-=t-a/2c$. In the following we do not consider
the last term occurring in \eqref{Omegax2}, since it is proportional to
$\delta(t)$ and thus only effective at the initial time $t=0$. We then
get from \eqref{Sij2}
\be
\lb{ddotx2}
\delta\ddot{x}_2(t)=\frac{2vcA}{a}\sqrt{\frac{2\pi}{b}}\theta(t)
\mathrm{Re}\left(e^{-\frac{t_-^2}{2b}-i\omega
  t_-}-e^{-\frac{t_+^2}{2b}-i\omega t_+}\right).
\ee
In a similar way one finds for the second time derivatives of the
other disturbances,
\bea
-M\delta \ddot{x}_1(t) &=& -\frac{\delta\Omega}{\delta x^1(t)}
=-\frac{Mv^2}{a}\theta(t)(\psi_+^{22}-\psi_-^{22})=\frac{v}{2c}\delta
\ddot{x}_2(t), \nonumber\\
-M\delta \ddot{x}_3(t) &=&-\frac{\delta\Omega}{\delta x^3(t)}=0.
\eea
One recognizes from these equations that $\delta x_1(t)$ is smaller
than $\delta x_2(t)$ by a factor $v/2c$; since we work in the
non-relativistic approximation $v/c\ll1$, we only have to consider
$\delta x_2(t)$ in the following. Therefore,
integrating \eqref{ddotx2} twice, one arrives at the result for $\delta
x_2(t)$,
\bea
\lb{resultx2}
\delta x_2(t)&=& \frac{2vcA\pi}{a}\sqrt{2b}
\ e^{-\frac{b\omega^2}{2}}\theta(t) \times\nonumber\\
& & \ \mathrm{Re}\left(\left[\frac{t}{\sqrt{2b}}+\sqrt{\frac{b}{2}}
\left(-\frac{a}{2cb}+i\omega\right)\right]\mathrm{erf}
\left[\frac{t}{\sqrt{2b}}+\sqrt{\frac{b}{2}}
\left(-\frac{a}{2cb}+i\omega\right)\right]\right.\nonumber\\
& & \  -\left[\frac{t}{\sqrt{2b}}+\sqrt{\frac{b}{2}}
\left(\frac{a}{2cb}+i\omega\right)\right]\mathrm{erf}
\left[\frac{t}{\sqrt{2b}}+\sqrt{\frac{b}{2}}
\left(\frac{a}{2cb}+i\omega\right)\right]\nonumber\\
& & \ \left. +\frac{1}{\sqrt{\pi}}\left(e^{-\left(\frac{t}{\sqrt{2b}}+
\sqrt{\frac{b}{2}}\left[-\frac{a}{2cb}+i\omega\right]\right)^2}-
e^{-\left(\frac{t}{\sqrt{2b}}+
\sqrt{\frac{b}{2}}\left[\frac{a}{2cb}+i\omega\right]\right)^2}\right)\right)
+C_1t+C_2,
\eea
where the constants $C_1$ and $C_2$ depend on the initial conditions.

We are interested in the limit where $t$ can be assumed large, that
is, after the wave has passed through and the measurement is
completed. Using the asymptotic property of the error function,
\ben
x\mathrm{erf}(x)+\frac{e^{-x^2}}{\sqrt{\pi}} \sim x,
\een
one gets
\be
\lb{asymptoticvalue}
\delta x_2(t) \sim -2vA\pi e^{-\frac{b\omega^2}{2}}+C_1t+C_2 \sim C_1t.
\ee
It is not surprising that asymptotically $\delta x_2(t)$ increases
linearly with time, since the dust particles comprising the cube do
not interact.

We want to impose initial conditions such that $\delta x_2(0)=0$ and
$\delta \dot{x}_2(0)=0$. After a straightforward calculation we find
that $C_2=0$ and
\be
\lb{C1}
C_1=-\frac{2vcA}{a\left[\left(\frac{a}{2bc}\right)^2+\omega^2\right]}
\sqrt{\frac{2\pi}{b}}
e^{-a^2/8bc^2}\left[\frac{a}{bc}\cos\left(\frac{\omega
      a}{2c}\right)-2\omega\sin\left(\frac{\omega a}{2c}\right)\right].
\ee
According to \eqref{asymptoticvalue}, the asymptotic behavior of
$\delta x_2(t)$ is determined by this constant $C_1$.

The next step in DeWitt's procedure is the application of
\eqref{SigmaIJ} in order to calculate the back reaction
on the gravitational wave, $\delta\psi_{\mu\nu}$. In this equation,
only the functional derivative of $\Omega$ (Eq. \eqref{interaction})
with respect to $\psi^{22}$ is non-zero, since
$T_{\mu\nu}\psi^{\mu\nu}=\rho\dot{x}_2^2\psi^{22}$; a derivative with
respect to $I$ and $J$ in \eqref{SigmaIJ} is thus a derivative with
respect to $\psi^{22}$. The left-hand side of \eqref{SigmaIJ} yields
\bdm
\frac{c^4}{4\pi G}\Box\delta\psi_{\mu\nu}(x).
\edm
On the right-hand side, the first term gives
\be
-\frac{\delta\Omega}{\delta\psi_{\mu\nu}}=-\rho\dot{x}_2^2\theta(t)
\delta^{\mu}_2\delta^{\nu}_2.
\ee
The second term also yields a contribution only for $\mu=2,\nu=2$, and
reads
\be
-\sum_i\frac{\delta\Omega}{\delta\psi_{22}\delta x^i}\delta x^i=
-v^2\left(2\rho D_1+v\frac{\partial\rho}{\partial x_2}D_0+
  v\frac{\partial\rho}{\partial x_2}D_1t\right)\theta(t),
\ee
where $D_0=-2A\pi e^{-b\omega^2/2}$ and $D_1=C_1/v$. Taking all
together, \eqref{SigmaIJ} reads
\be
\Box\delta\psi_{22}=\frac{4\pi G}{c^4}v^2\left(2\rho
  D_1+v\frac{\partial\rho}{\partial x_2}D_0+
  v\frac{\partial\rho}{\partial x_2}D_1t\right)\theta(t).
\ee
We solve this equation with the help of the usual retarded Green function,
\bdm
D_r(t-t',\mathbf{r}-\mathbf{r'})=\frac{\theta(t-t')}
{4\pi\vert\mathbf{r}-\mathbf{r'}\vert}\delta\left(t-t'-\frac{\vert\mathbf{r}-\mathbf{r'}\vert}{c}\right) ,
\edm
under the assumption that
$\vert\mathbf{r}-\mathbf{r'}\vert\approx\vert\mathbf{r}\vert\equiv r$; we
also introduce the retarded time
\bdm
t_r\equiv t-\frac{\vert\mathbf{r}-\mathbf{r'}\vert}{c}\approx
t-\frac{r}{c}.
\edm
After some straightforward calculations, we get the result
\be
\lb{deltapsi22}
\delta\psi_{22}(\mathbf{r},t)=\frac{GMv^2}{c^4r}
\left(1-\frac{2C_1}{v}\right)\theta(t_r).
\ee
The first term on the right-hand side corresponds to the first term on
the right-hand side of \eqref{SigmaIJ}; the form of the resulting term
is well-known from the generation of gravitational waves by, for
example, the circular motion of two masses (here it is a consequence
of the interaction \eqref{interaction}).
 The second term originates
from the second term in \eqref{SigmaIJ} and describes the back
reaction coming from $\delta x^i$. We expect it to be much smaller
than the first term. Let us perform some numerical estimates.

We take, for example, the values $\omega=1$ Hz and $a=1$ m (but the
precise values are not important). We then
have
\bdm
\frac{\omega a}{2c}\sim 10^{-9}\ll 1,
\edm
and we can therefore approximate $\cos\frac{\omega a}{2c}\approx 1$
and $\sin\frac{\omega a}{2c}\approx\frac{\omega a}{2c}$ in
\eqref{C1}. The ratio of the second term to the first term in the
parentheses of \eqref{C1} is then given by $\omega^2b$, which is much
bigger than 1 if we assume a narrow packet for the gravitational wave
\eqref{PSI} (as we shall do). We then also find that the first term in
the denominator of \eqref{C1} is much smaller than the second term.
We thus get
\be
\lb{2C1v}
\frac{2C_1}{v}\approx 4A\sqrt{\frac{2\pi}{b}}e^{-a^2/8bc^2}.
\ee
The size of the back reaction in \eqref{deltapsi22} is thus
proportional to the amplitude of the gravitational wave as
expected. If we take $A/\sqrt{b}\sim 10^{-20}$ (which should be a
realistic value for an astrophysical gravitational wave), we see that
the back-reaction term is tiny. Since the exponential in \eqref{2C1v}
is $\approx 1$ for our values of parameters, we get from
\eqref{deltapsi22},
\be
\delta\psi_{22}(\mathbf{r},t)\approx\frac{GMv^2}{c^4r}
\left(1-4A\sqrt{\frac{2\pi}{b}}\right).
\ee

We have obtained this result from the application of the general
expression \eqref{SigmaIJ}. This formula follows after neglecting
certain terms in the more general formula \eqref{exp.of ap.}. But can
this be justified? After all, a gravitational wave is not a massive
apparatus in the original sense of DeWitt's analysis.
If we compare \eqref{SigmaIJ} with
\eqref{exp.of ap.}, there are only two terms that are being
neglected, the third and fifth terms of \eqref{exp.of ap.}.
One then needs to justify
neglecting these two terms. In our case, the fifth term is identically zero
because the interaction term \eqref{interaction} is linear in the
field $\psi^{\mu\nu}$ and
therefore its second functional derivative is zero. To justify neglecting
the third term, notice that this term is proportional to the square of
$\delta\psi_{22}$.
But, according to the approximate calculation that has been carried
out, see \eqref{deltapsi22}, $\delta\psi_{22}$
 is of order $v^2/c^4$. One can therefore assume
that the third term must be of order $v^4/c^8$, and therefore it is safe to
neglect it in the non-relativistic approximation. The approximation
\eqref{SigmaIJ} is thus a good one because of the form of the
interaction and the non-relativistic approximation.

We now need to consider a suitable `apparatus quantity' of the
gravitational wave to be measured, cf. \eqref{deltaA}.
 We choose the energy density of the gravitational wave,
\be
A(\Psi_{\mu\nu})\equiv T^{\rm GW}_{00}=\frac{c^2}{8\pi G}
  \dot{\Psi}_{\mu\nu}\dot{\Psi}^{\mu\nu}\ .
\ee
The change of $T^{\rm GW}_{00}$ is given by
\be
\delta T^{\rm GW}_{00}=\frac{c^2}{8\pi G}\int
d^3x'dt'\frac{\delta T^{\rm GW
  }_{00}}{\delta\Psi^{\mu\nu}}\delta\Psi^{\mu\nu}
= -\frac{c^2}{4\pi G}\ddot{\psi}^{22}\delta\psi_{22} , \lb{obs.dist.}
\ee
since $\delta\dot{\psi}_{22}=0$ (neglecting, again, a delta-function
contribution).
If we use again $b^{-1}\ll\omega^2$ and set $t\approx x_1/c$ (because
only then do we get a noticeable contribution from a packet peaked
around $t-x_1/c$), we get $\ddot{\psi}^{22}\approx
-A\omega^2\sqrt{2\pi/b}$, and therefore
\be
\delta T^{\rm GW}_{00}\approx \frac{A\omega^2Mv^2}{4\pi
  rc^2}\sqrt{\frac{2\pi}{b}} \left(1-4A\sqrt{\frac{2\pi}{b}}\right).
\ee
(Note that the dependence on $a$ has dropped out in this limit.)
The measurement of $\delta T^{\rm GW}_{00}$ thus yields directly
information about the kinetic energy of the cube (and also about $a$
if higher-order corrections are taken into account).

These results, although derived for a classical mass cube, are also
valid, at least approximately, for a quantized system. This follows
from the fact that the equations of motion for the center of mass of
the free quantum system are also solved by (\ref{free system}). The
solutions of the uncoupled equations are therefore the same; however,
in the case of a quantum system, the interaction term may differ from
(\ref{interaction}) by terms of order $\hbar$, leading to corresponding
quantum corrections to the disturbances. These corrections should be
negligible if the quantum system is large enough, in which case our
solutions are also valid for a gravitational wave packet interacting
with a quantum mass cube. The effect of the quantum uncertainties of
the system on the disturbance of the apparatus observable is estimated
below.

Regarding the EH gedanken experiment, it is of interest to consider
the limit of $\delta T^{\rm GW}_{00}$ in which the amplitude $A$ tends
to zero. Taking $A\rightarrow0$ leads to a
vanishing disturbance of the apparatus observable:
\begin{eqnarray*}
\lim_{A\rightarrow0}\delta T^{\rm GW}_{00}=0\ ,
\end{eqnarray*}
which suggests that obviously no measurement can be achieved within this
limit. Another way of coming into contact with the EH gedanken experiment, at
least on a heuristical level, is to introduce quantum uncertainties
for the mass-cube. The particles comprising the cube obey
the uncertainty principle. These uncertainties limit the
accuracy for the measurement of the edge length $a$ and introduce
an uncertainty that depends on the particles' position uncertainty
$\Delta x$. A simultaneous momentum measurement will be restricted by
the quantum mechanical momentum uncertainty $\Delta
p \gtrsim{\hbar}/{\Delta x}$. It is only at this stage that the
quantum theory comes into play (and only in the weak form of the
uncertainty relation). The resulting classical uncertainty for
$T_{00}^{\rm GW}$ then becomes
\begin{eqnarray*}
\Delta T_{00}^{\rm GW}&=&\sqrt{\left(\frac{\partial \delta
      T_{00}^{\rm GW}}{\partial a}\Delta
    a\right)^2+\left(\frac{\partial\delta T_{00}^{\rm GW}}{\partial
      v}\Delta v\right)^2}\nonumber\\
&\ge&\sqrt{\left(\frac{\partial \delta T_{00}^{\rm GW}}{\partial a}\Delta
    x\right)^2+\left(\frac{\partial \delta T_{00}^{\rm GW}}{\partial
      v}\frac{\hbar}{2m\Delta x}\right)^2}\label{heuristic unc.} \ .
\end{eqnarray*}
The minimum uncertainty is found for
\begin{equation*}
\Delta x_{\rm min}= \sqrt{\frac{\hbar}{2m}\frac{\delta T_{00}^{\rm
    GW},_v}{\delta T_{00}^{\rm GW},_a}} \ ,
\end{equation*}
and therefore
\begin{eqnarray*}
\Delta T_{00}^{\rm GW}\ge\sqrt{\frac{\hbar}{m}\delta
  T_{00}^{\rm GW},_v \delta T_{00}^{\rm GW},_a}\label{uncertaintoo}\ .
\end{eqnarray*}
Thus, a coupling between a weak gravitational wave and a test body
obeying the uncertainty principle in the manner described above
unavoidably transfers a minimum amount of uncertainty to the classical
system, as can be seen here for the averaged energy density of the
scattered gravitational wave (for more arguments on
the transfer of uncertainties, see Sec. \ref{cl-quant.systems}). But
this disagrees with the requirement in \cite{EH} that the
gravitational wave, considered as a classical system, should be free
of all uncertainties during the whole measurement. This indicates
another weakness of the arguments presented in the EH
gedanken experiment. We thus conclude from this discussion that the
corresponding assumption made in \cite{EH} is unrealistic. Whether the
derived minimum uncertainty is big enough to strictly invalidate the
EH-arguments is a different question and beyond the scope of our
discussion.


\section{Interacting classical-quantum systems}\lb{cl-quant.systems}

In this section, we construct an explicit model to describe the
interaction of a classical gravitational wave with a quantized
field. This provides a counterexample to the claims made in
\cite{DeWitt62} and \cite{EH} that a system coupled to a quantum
system must, for consistency reasons, also be of quantum nature.
A number of different methods have been proposed to model
`mixed' classical-quantum systems; here, we use the formalism of
ensembles in configuration space \cite{HR}. This formalism is
applicable to both classical and quantum systems and allows a general
and consistent description of interactions between them. In
particular, the correct equations of motion for the classical and
quantum sectors are recovered in the limit of no interaction,
conservation of probability and energy are satisfied, uncertainty
relations hold for conjugate quantum variables, and the formalism
allows a back reaction of the quantum system on the classical system.

We consider one of the simplest models: a two-dimensional version of
the scalar theory of gravity of Nordstr\"{o}m
\cite{Nordstroem13,straumann} with
a quantized massive scalar field. Although not a viable theory of
gravity \cite{OUP,Harvey65}, the theory of Nordstr\"{o}m
appears to be the simplest one that has all the ingredients that are
necessary to model a gedanken experiment of the Eppley and Hannah
type. There are gravitational waves in the theory, the coupling to the
scalar field is uniquely determined, and the calculations simplify
somewhat because the whole theory can be formulated in Minkowski
space--time (although the reformulation due to Einstein and Fokker
\cite{EinsteinFokker14} shows that one may relax the requirement of a
flat space--time and provide a geometric interpretation of the
theory). The simple model discussed in this section already shows the
main features expected of a more realistic description while
presenting fewer technical complications, the main advantage being
that the interaction term is simpler than the one derived from general
relativity. This will allow us to get explicit expressions for the
state of the coupled classical-quantum system that are valid to first
order in the coupling parameter for particular classes of fields.

The scalar theory of gravity of Nordstr\"{o}m is based on the
Lagrangian density \cite{OUP}
\begin{equation}
\mathcal{L}=\mathcal{L}_{N}+\mathcal{L}_{\rm matter}=-\frac{1}{2}\eta ^{\mu \nu
}\phi ,_{\mu }\phi ,_{\upsilon }-gT\phi +\mathcal{L}_{\rm matter}, \nonumber \\
\end{equation}
where $\eta _{\mu \nu }={\rm diag}(-1,1)$ is the two-dimensional
Minkowski metric,
the commas indicate partial derivatives, $g$ is the coupling constant,
$T=$ $\eta ^{\mu \nu }T_{\mu \nu }$ is the trace of the energy
momentum tensor $T_{\mu \nu }$ and $\mathcal{L}_{\rm matter}$ is the
Lagrangian density of matter. It will be convenient to consider the
Hamiltonian formulation of the theory, which in two dimensions and for
the particular case where the matter consists of a massive
scalar field $\psi$ takes the form
\begin{equation}
H = \frac{1}{2}\int dx\,\left( \pi _{\phi }^{2}+\phi ^{\prime 2}\right) +%
\frac{1}{2}\int dx\,\left( \pi _{\psi }^{2}+\psi ^{\prime 2}+m^{2}\psi
^{2}\right) +gm^{2}\,\int dx\,\phi \psi ^{2} \label{HNsgMsf}. \\
\end{equation}
Here, the prime indicates a derivative with respect to the spatial coordinate.

The description of `mixed' classical-quantum systems of reference
\cite{HR} is based on a canonical formalism for describing statistical
ensembles on configuration space (which is here the space spanned by
the fields $\phi$ and $\psi$).
The state of the system is described
in terms of a probability $P$ together with its canonically conjugate
variable $S$, and the equations of motion are derived from an ensemble
Hamiltonian. For the model that we are considering in this section,
the equations for $P$ and $S$ are of the form
\begin{eqnarray}
&&\dot{S}+\int dx\left\{ \frac{1}{2}\left[ \left( \frac{\delta S}{\delta
\phi }\right) ^{2}+\left( \frac{\delta S}{\delta \psi }\right) ^{2}\right] +%
\frac{\hbar ^{2}}{8}\left[ \frac{1}{P^{2}}\left( \frac{\delta P}{\delta \psi
}\right) ^{2}-\frac{2}{P}\frac{\delta ^{2}P}{\delta \psi ^{2}}\right]
\right.   \label{Sdot} \\
&&\qquad \qquad \qquad \qquad \left. -\frac{1}{2}\phi \phi ^{\prime \prime }+%
\frac{1}{2}\psi \left[ -\psi ^{\prime \prime }+\left( 1+2g\phi \right)
m^{2}\psi \right] \right\} =0  \nonumber
\end{eqnarray}
and
\begin{equation}
\dot{P}+\int dx\left[ \frac{\delta }{\delta \phi }\left( P\frac{\delta S}{%
\delta \phi }\right) +\frac{\delta }{\delta \psi }\left( P\frac{\delta S}{%
\delta \psi }\right) \right] =0,  \label{Pdot}
\end{equation}
where the overdot indicates a derivative with respect to the time coordinate.

Equation (\ref{Sdot}) has the form of a modified Hamilton--Jacobi
equation, and in the limit where $\hbar$ goes to zero it becomes
identical to the Hamilton--Jacobi equation that corresponds to equation
(\ref{HNsgMsf}). Equation (\ref{Pdot}) can be interpreted as a
continuity equation for the probability $P$. There are some subtle
issues concerning the physical interpretation of $S$ within the
formalism of ensembles in configuration space which we can discuss
only briefly here. To maintain full generality, $S$ should not be
regarded as a field momentum density potential. In particular, for
an ensemble of classical fields with uncertainty described by
probability $P$, it will not be assumed that the field momentum
density of a member of the ensemble is a well-defined quantity
proportional to the functional derivative of $S$, as it is done in
the usual deterministic interpretation of the Hamilton--Jacobi
functional equation. This avoids forcing a similar deterministic
interpretation in the quantum and quantum-classical cases. A
deterministic picture can be recovered for classical ensembles
precisely in those cases in which trajectories are operationally
defined \cite{HR}.


\subsection{Solution for the non-interacting case}\lb{CQ-non-interacting}

We first derive solutions for the non-interacting case (i.e., we set
$g=0$). We assume
\begin{eqnarray}
S \left[ \phi,\psi,t \right)&=& S^{c} \left[ \phi,t \right) - E^{q} t
, \label{S-WLP} \\
P \left[ \phi,\psi,t \right)&=& P^{c}\left[ \phi,t \right) P^{q}\left[
  \psi \right],    \label{PWLP}
\end{eqnarray}
where $E^{q}$ is a constant. With this ansatz, (\ref{Sdot})
and (\ref{Pdot}) take the simpler form
\begin{eqnarray}
-\dot{S^c}+E^q&=&\frac{1}{2}\int dx\left\{ \left( \frac{\delta
      S^c}{\delta \phi _{x}}%
\right) ^{2}-\frac{\hbar ^{2}}{A^{q}}\frac{\delta
^{2}A^{q}}{\delta \psi _{x}^{2}} -\phi _{x}\phi _{x}^{\prime
\prime }-\psi _{x}\left[ \psi _{x}^{\prime \prime }- m^{2}\psi
_{x}\right] \right\}, \label{SdotWLP} \\
\dot{P^c}&=&-\int dx\left[ \frac{\delta }{\delta \phi _{x}}\left(
    P^c\frac{\delta S^c}{\delta \phi _{x}}\right) \right] =0,
\label{PdotWLP}
\end{eqnarray}
where we have introduced $A^{q } \equiv \sqrt{P^{q }}$. To solve these
equations, we will use standard techniques developed for the
Schr\"{o}dinger functional representation of quantum field theory
\cite{Jackiw90} \cite{LongShore96}. Here, however, we work in a
representation where the pair of canonically conjugate functionals $P$
and $S$ are taken as fundamental variables. Although it is possible to
introduce a wave functional for the total system by means of a
transformation of the form $\Psi=\sqrt{P} e^{iS/\hbar}$, there is no
clear advantage in doing this as (\ref{SdotWLP}) and
(\ref{PdotWLP}) do not become linear in this case.

We consider solutions which are of the form
\begin{eqnarray}
S^{c} \left[ \phi,t \right)&=&\frac{1}{2}\int \int dydz\;\phi
_{y}F_{yz}(t)\phi _{z},
\nonumber\\
P^{c}\left[ \phi,t \right)&=&N^{c}(t)\;e^{ -\frac{1}{2}\int \int
dadb\;\left( \phi _{a}-\beta _{a}(t) \right) K_{ab}(t)\left( \phi _{b}-\beta _{a}(t) \right) } ,
\nonumber\\
P^{q}\left[ \psi \right] &=& N^{q}\left( \frac{1}{\hbar }\int
\int dydz\;\gamma _{y}G_{yz}\psi _{z}\right) ^{2}
e^{-\frac{1}{\hbar }\int \int dadb\;\psi_{a}G_{ab}\psi _{b}} .
\nonumber
\end{eqnarray}
Equation (\ref{SdotWLP}) leads to
\begin{eqnarray}
E^{q } -\frac{\hbar }{2}\int dxG_{xx}-\frac{\int \int \int dxdydz\;\gamma
_{y}G_{yx}G_{xz}\psi _{z}}{\int \int dydz\;\gamma _{y}G_{yz}\psi _{z}} &=&0, \label{EqEgamma}\\
\dot{F}_{yz}+\int dx\;F_{yx}F_{xz}-\partial _{z}^{2}\delta (y-z)&=&0, \label{EqKernelF}\\
-\int dx\;G_{yx}G_{xz} -\left[ \partial _{z}^{2}- m^{2}\right] \delta (y-z)&=&0,  \label{EqKernelG}
\end{eqnarray}
while equation (\ref{PdotWLP}) leads to
\begin{eqnarray}
\frac{\dot{N}^{c }}{N^{c }} -\frac{1}{2}\int \int dydx\;\left( \beta
_{y}K_{yx}\,\beta_{x} \right)^{.} +\int dx\;F_{xx}\, &=&0, \label{EqN}\\
\int dy\;\left( \beta _{y}K_{yz}\right) ^{\cdot }\,+\int \int dydx\;\beta
_{y}K_{yx}\,F_{xz} &=&0,  \label{Eqbeta}\\
-\frac{1}{2}\dot{K}_{yz}-\int dx\;K_{yx}F_{xz} &=&0 \label{EqKernelK}.
\end{eqnarray}%

To solve these equations, we introduce (real) basis functions
$f_{x}^{(k)}$ that satisfy the eigenvalue equation
\begin{equation}
-\partial _{x}^{2}f_{x}^{(k)}=k^{2}f_{x}^{(k)} \nonumber
\end{equation}
as well as orthonormality and completeness relations of the form $\int
dx\;f_{x}^{(k)}f_{x}^{(m)} =\delta _{km}$ and
$\sum_{k}f_{x}^{(k)}f_{y}^{(k)} =\delta \left( x-y\right)$. We have
assumed discrete eigenvalues to simplify the notation, but the
extension to continuous ones is straightforward. We now expand all
quantities in terms of these basis functions and use these expressions
in (\ref{EqEgamma})--(\ref{EqKernelK}). Equations
(\ref{EqKernelF}), (\ref{EqKernelG}) and (\ref{EqKernelK}) lead to
representations for the kernels of the form
\begin{eqnarray}
F_{xy}&=&-\sum_{k}k\tan \left( kt\right) f_{x}^{(k)}f_{y}^{(k)}, \label{Fxy}\\
G_{xy}&=&\sum_{k}\sqrt{k^{2}+m^{2}}f_{x}^{(k)}f_{y}^{(k)}, \label{Gxy}\\
K_{yx}&=&\sum_{k}\frac{\tau_k}{ \cos ^{2} \left( kt\right)
}f_{x}^{(k)}f_{y}^{(k)}, \label{Kxy}
\end{eqnarray}
where the $\tau_k$ are arbitrary constants. Equations (\ref{EqEgamma})
and (\ref{Eqbeta}) lead to representations for the functions $\gamma
_{x}$ and $\beta _{x}$ and also determine the value of $E^q$,
\begin{eqnarray}
\gamma _{x}&=&f_{x}^{(a)}, \label{gamma} \\
E^{q } &=&\frac{\hbar }{2}\int dx\;G_{xx}+ \sqrt{a^{2}+m^{2}}, \label{Eq}\\
\beta _{x}&=&\sum_{k} w_k \cos(kt) f_{x}^{(k)} \label{beta},
\end{eqnarray}
where the $w_k$ and $a$ are arbitrary constants. The term on the
right-hand side of (\ref{Eq}) diverges and needs to be
renormalized. Fortunately, the divergence can be easily isolated and
removed in this formalism: it is present in the first term of the
right-hand side. This prescription for dealing with the energy
renormalization  is essentially the same one that is used when dealing
with the Schr\"{o}dinger wave functional representation of quantum
field theory \cite{LongShore96}. In our case, the issue of the
renormalization prescription is not a crucial one and we will not
discuss it further. Finally, $N^{q }$ is proportional to a constant
while $N^{c }$ is determined by equation (\ref{EqN}) and is formally
given by
\begin{equation}
N^{c } \sim  \frac {1} { {\Pi_{k}\cos \left( kt\right) }} \label{Nc}
\end{equation}
in the generic case where none of the $\tau_k$ vanishes.

The physical interpretation of these solutions can be summarized
briefly as follows. The solution \{$S^{c} \left[ \phi,t \right)$, $P
\left[ \phi,\psi,t \right)$\} provides the field-theoretic
generalization of a solution which describes a particular ensemble in
configuration space for a one-dimensional classical harmonic oscillator
(i.e., a one-dimensional oscillator state which is prepared with zero
momentum and no momentum uncertainty, but localized in space -- see the
Appendix for details); $A^{q}\left[ \psi \right]$ is the Schr\"{o}dinger
wave functional for a one-particle state specified by the eigenfunction
$\gamma _{x}=f_{x}^{(a)}$ and of energy $E^q$ \cite{LongShore96}.


\subsection{Solution for the interacting case ($g \ll 1$)}\lb{CQ-interacting}

We now want to turn on the interaction and consider the case where $g
\ll 1$. A general solution to (\ref{Sdot}) and (\ref{Pdot})
valid to first order in $g$ seems quite complicated, although in
principle possible. We derive here a solution that is correct to first
order in $g$ for a particular class of states. In this solution, the
expression for $S$, given by (\ref{S-WLP}), remains of the
same form, but the expression for $P$, given by (\ref{PWLP}),
is modified to $P \left[ \phi,\psi,t \right) = P^{c}\left[ \phi,t
\right) P^{cq}\left[ \phi,\psi \right]$ with
\begin{equation}
P^{cq}\left[ \phi,\psi \right] = N^{q}[\phi]\left( \frac{1}{\hbar }\int
\int dydz\;\gamma _{y}G_{yz}[\phi] \psi _{z}\right) ^{2}
e^{-\frac{1}{\hbar }\int \int dadb\;\psi
_{a}G_{ab}[\phi] \psi _{b}} .
\nonumber
\end{equation}
That is, we now allow the kernel $G_{xy}$ and the normalization factor
$N^q$ to be functionals of $\phi$. With this new ansatz,
(\ref{Sdot}) and (\ref{Pdot}) lead to a set of equations similar to
the set (\ref{EqEgamma})-(\ref{EqKernelK}) that we
derived previously. Note that we do not distinguish explicitly between
quantities that are evaluated with $g = 0$ and with $g \neq 0$ in the
rest of this section in order not to clutter the notation.

We first consider the three equations derived from
(\ref{Sdot}): equations (\ref{EqEgamma}) and (\ref{EqKernelF}) remain
as before, but (\ref{EqKernelG}) is replaced by
\begin{equation}
-\int dx\;G_{yx}G_{xz} -\left[ \partial _{z}^{2}- \left( 1+2 g \phi_z
  \right) m^{2}\right] \delta (y-z) = 0.  \label{EqKernelGg}
\end{equation}
A solution of (\ref{EqKernelGg}) that is correct to first
order in $g$ is given by
\begin{eqnarray}
G_{xy}[\phi]&=&\sum_{k}\sqrt{k^{2}+m^{2}\left(1+2 g I^k
  \right)}\;g_{x}^{(k)}g_{y}^{(k)} \nonumber\\
&\simeq& \sum_{k}\sqrt{k^{2}+m^{2}} \left( 1+ \frac{g
    I^k}{k^{2}/m^{2}+1} \right)\, g_{x}^{(k)}g_{y}^{(k)},
\nonumber
\end{eqnarray}
where the dependence of $G_{xy}[\phi]$ on $\phi_x$ comes in through
\begin{equation}
I^k=\int dx\;f_{x}^{(k)} \phi_x f_{x}^{(k)}. \nonumber
\end{equation}
The $g_{x}^{(k)}$ are modified eigenfunctions determined by first
order perturbation theory,
\begin{equation}
g_{x}^{(k)}=f_{x}^{(k)}+\sum_{n \neq k} \epsilon^{(kn)} f_{x}^{(n)} \nonumber
\end{equation}
with
\begin{equation}
\epsilon ^{(kn)}=2gm^{2}\sum_{n\neq k}\left[ \frac{\int dx\;f_{x}^{(k)}\phi
_{x}f_{x}^{(n)}}{k^{2}-n^{2}}\right] \ll 1. \nonumber
\end{equation}
Equation (\ref{EqKernelF}) is independent of both $g$ and $G_{xy}$ and
therefore the previous solution for $F_{xy}$, equation (\ref{Fxy}),
remains valid. To solve (\ref{EqEgamma}), we require a
$\gamma_x$ that satisfies the integral equation $\int dx\; \gamma_x
G_{xy} = \lambda \gamma_y$ where $\lambda$ is a constant. The
expression
\begin{equation}
\gamma_{x}=f_{x}^{(a)}+\sum_{n \neq a} \epsilon^{(an)} f_{x}^{(n)}
\nonumber
\end{equation}
is correct to first order in $g$. We also need to ensure that $E^q$
remains a constant (i.e., independent of $\phi_x$). This requirement is
satisfied if we restrict to values of $a$ that are large enough to
have $a^2/m^2 \sim 1/g$. In this case, the correction to the quantum
energy $E^q$ becomes negligible (of order $g^2$) and can be ignored,
since
\begin{eqnarray}
E^{q } &=& \frac{\hbar }{2}\int dx\;G_{xx}+ \sqrt{a^{2}+m^{2}} \left(
  1+ \frac{g I^a}{a^{2}/m^{2}+1} \right) \nonumber\\
& \simeq & \frac{\hbar }{2}\int dx\;G_{xx}+ \sqrt{a^{2}+m^{2}}. \nonumber
\end{eqnarray}
Therefore, our solution is correct to first order in $g$ provided the
quantized field is in a state of high enough energy.

We now consider the three equations derived from
(\ref{Pdot}): equations (\ref{EqN}) and (\ref{EqKernelK}) remain as
before, but (\ref{Eqbeta}) is replaced by
\begin{eqnarray}
\int dy\;\left( \beta _{y}K_{yz}\right) ^{\cdot }\,&+& \int \int
dydx\;\beta_{y}K_{yx}\,F_{xz} + \frac{\int \int dydx\;\frac{\delta
    N^q}{\delta G_{yx}}\frac{\delta G_{yx}}{\delta \phi_z}}{N^q}
\nonumber\\
&-& \frac{ 2 \int \int dydx\;\gamma _{y} \frac {\delta G_{yx}}{\delta
    \phi_z}\psi_{x}}{ \int\int dydx\;\gamma _{y}G_{yx}\psi _{x}}
+ \frac {1}{\hbar} \int \int dydx\;\psi _{y} \frac {\delta
  G_{yx}}{\delta \phi_z}\psi _{x} = 0 .
\label{Eqbetag}
\end{eqnarray}
The last three terms in (\ref{Eqbetag}) are all of order $g$,
and the correction due to these terms becomes negligible (of order
$g^2$) if we restrict to gravitational waves where the constants
$\tau_k \sim 1/g$ in equation (\ref{Kxy}). Under this assumption, the
expressions for $K_{xy}$, $\beta_x$ and $N^c$ given by
(\ref{Kxy}), (\ref{beta}) and (\ref{Nc}) all remain valid. Therefore,
our solution is correct to first order in $g$ provided we restrict to
gravitational waves $\phi_x$ that are sharply peaked about $\beta_x$.


\subsection{Discussion}\lb{CQ-discussion}

We have considered a classical gravitational wave interacting with a
quantum field in the context of a two-dimensional model and we have
derived an explicit solution for this `mixed' classical-quantum
system. Our solution is valid to first order in the coupling parameter
$g$ and for a particular class of states. While the argument of Eppley
and Hannah is based on supposed inconsistencies that would arise when
trying to couple a classical gravitational wave to a quantum system,
the results of this section suggest that there is no fundamental
principle that excludes such systems. This is further evidence that
the argument for quantizing gravity cannot be based on the claim that
the non-quantization of gravity would lead to logical inconsistencies
of this sort. It is of interest to discuss briefly one particular
aspect of the solution derived here. In the non-interacting case, the
probability $P[\phi, \psi, t)$ of the total system is the product of
the probabilities of each of the subsystems, as expressed by
(\ref{PWLP}), and this means that the uncertainty associated with the
quantized scalar field cannot affect any averages that we calculate
for observables of the classical gravitational wave (in particular,
estimates of uncertainties such as the root mean square deviation will
be independent of the state of the quantized scalar field). But once
the interaction is turned on, and in particular as we incorporate
corrections that are of higher order in $g$, the uncertainty of the
observables of the classical gravitational wave will depend on the
particular state of the quantized scalar field since the probability
of the total system will be a rather complicated function of the
fields which will no longer factor into a simple product. This
confirms one of the results obtained in the previous section, that
uncertainty is transferred from the quantum system to the classical
system as a result of the interaction.
Since the model of interacting classical-quantum systems discussed in
this section provides an  explicit counterexample to DeWitt's claim
that ``...the quantization of a given system implies also the
quantization of any other system to which it can be coupled [...]
therefore, the quantum theory must immediately be extended to all
physical systems, including the gravitational field'' \cite{DeWitt62},
it will be useful to examine DeWitt's argument to clarify this issue.


\section{Critique of DeWitt's argument}

In order to explain and present DeWitt's argumentation, it is necessary
to introduce an appropriate concept used by him: the
Peierls brackets. The Peierls brackets are a generalization of the
ordinary Poisson brackets. Generalization because they directly follow
from the action of the considered system and do not require the
canonical coordinates and momenta to be defined in advance. But
they are nevertheless identical with the usual Poisson brackets for
standard canonical systems (if it is possible to arrive
  at an unconstrained Hamiltonian formalism corresponding to the
  initial Lagrangian formalism, which is the case for non-singular
  Langrange functions). This fact allows one to apply the
correspondence principle, that is, to introduce commutators, to the
measurement analysis, which in turn enabled DeWitt to obtain his final
result \eqref{result, dewitt2} below.

To explain the meaning of the Peierls brackets, it is convenient to
recall the formalism of Sec. \ref{DeWitt m.a.}. Consider a
system $S[\phi]$. Its equation of motion may be solved by $\phi_0$. A
change of the action functional $S$ will in general lead to an
equation of motion whose solution deviates from $\phi_0$. Under the
requirement that this change is small, it is natural to
expand the action functional around the free solution. Omitting higher
terms allows one to calculate the deviation $\delta \phi$ of the
system variable. The formal solution via the Green-function method
reads
\begin{equation*}
\delta \phi^j=gG^{ij}\Omega,_i \ ,
\end{equation*}
where $G^{ij}$ denotes the Green function in the condensed notation.
The additional action term not only produces disturbances within the
dynamical variable $\phi_0$, but also in any observable $B$ built out
of these. The expansion of $B$ up to first order in $g$ leads to
\begin{equation*}
B[\phi_0+\delta\phi]\approx B[\phi_0]+B,_i\delta\phi^i \ ,
\end{equation*}
where $B,_i\delta\phi^i$ is formally given by
\begin{equation*}
B,_i\delta\phi^i=gB,_iG^{ij}\Omega,_j\equiv\delta_{\Omega}B \ .
\end{equation*}
Before we introduce the Peierls bracket, it is convenient to
define the operation
\begin{equation*}
D_{\Omega}B\equiv\lim_{g\rightarrow0}\frac{1}{g}\delta_{\Omega}B=B,_iG^{ij}\Omega,_j
\ .
\end{equation*}
To comply with condition \eqref{ret.condition}, one can choose the
retarded Green function for the calculation of $\delta\phi$. Doing
this leads to the interpretation of $D_{\Omega}B$ as the retarded
change which $\Omega$ causes within $B$.

The Peierls brackets are now defined as the difference between the
effects which two quantities, say $A$ and $B$, have on each other
in the sense described above,
\begin{eqnarray*}
(A,B)\equiv D_AB-D_BA \ .
\end{eqnarray*}
A formal calculation for the small disturbance of the apparatus
variable $\theta$ yields
\begin{eqnarray*}
\delta\theta^J&=&gG^{IJ}\left(\Omega,_I+\Omega,_{Ii}\delta\phi^i\right) \
\end{eqnarray*}
as a solution of \eqref{SigmaIJ}. To approximate the disturbance of
any \emph{apparatus} observable $A$, one can use a procedure analogous
to the one used above for system observables $B$; that is, approximate
$A[\theta]\approx A[\theta_0]+\delta A$ with
\begin{eqnarray}
\delta A&=&
A,_I\delta\theta^I=gA,_IG^{IJ}\left(\Omega,_J+\Omega,_{Ji}\delta\phi^i\right)\nonumber\\
&=&gA,_IG^{IJ}\Omega,_J+gA,_IG^{IJ}\Omega,_{Ji}\delta\phi^i\nonumber\\
&=&gD_{\Omega}A\quad\quad\;\;+\quad\quad
g^2A,_IG^{IJ}\Omega,_{Ji}G^{ij}\Omega,_j\nonumber\\
&=&gD_{\Omega}A\quad\quad\;\;+\quad\quad
g^2(D_{\Omega}A),_i G^{ij}\Omega,_j\nonumber\\
&\approx&gD_{\Omega}A\quad\quad\;\;+\quad\;\quad g^2
D_{\Omega}\left(D_{\Omega}A\right)\ ,\label{5.row}
\end{eqnarray}
where the omission of the $(D_{\Omega}A),_I G^{IJ}\Omega,_J$ term in the
last line is justified by choosing the apparatus to be `macroscopic' compared to the
system (see Sec. \ref{DeWitt m.a.}).

Given a particular choice of apparatus observable, one would like to use
the expression for $\delta A$ to express a system observable in terms
of the `experimental data'. Since this is not possible in general, DeWitt
considered the following special case. If the coupling term $\Omega$ is
chosen so that it satisfies
\begin{equation}
D_{\Omega}A=s\label{D one obs}
\end{equation}
for a particular system observable $s$, \eqref{5.row} becomes
\begin{eqnarray*}
\delta A=gs + g^2 D_{\Omega}s \ .\label{securecond}
\end{eqnarray*}
Then, if $D_{\Omega}s$ only depends on apparatus variables, $s$ can be
expressed in terms of experimental data,
\begin{equation}
s=\frac{\delta A}{g}-gD_{\Omega}s \ . \label{sexp}
\end{equation}
Since $s$ is a function of the experimental data, there will be an uncertainty
associated with $s$ that reflects limited knowledge of the apparatus
quantities. If, say, $A$ and $D_{\Omega}s$ are only known up to $\Delta A$
and $\Delta D_{\Omega}s$, the uncertainties will propagate as
\begin{equation*}
\Delta s^2=\frac{\Delta A^2}{g^2}+g^2 \left(\Delta D_{\Omega}s\right)^2\ .
\end{equation*}
If this is minimized with respect to the coupling constant $g$, it becomes
\begin{equation}
\Delta s= \sqrt{2\Delta A \Delta D_{\Omega}s}\ .\label{Delta s}
\end{equation}
This equation is not very interesting if one remains at the classical
level, because both $\Delta A$ and $\Delta D_{\Omega}s$ can be made
arbitrarily small and hence $s$ can be measured with arbitrary
accuracy. But if one refers to quantum
mechanics, limitations for the product of uncertainties of conjugate
observables arise. DeWitt noticed how to reformulate
\eqref{Delta s} to contain exactly such a product; that is, to choose
\begin{equation}
\Omega =sC \ ,\label {action one obs}
\end{equation}
where $C$ is an apparatus variable conjugate to
$A$, that is, with $(A,C)=1$. Using the identity
\linebreak$D_{r(\phi)s(\theta)}t(\phi)=s(\theta)\left(D_{r(\phi)}t(\phi)\right)$
enables \eqref{Delta s} to be rewritten as
\begin{eqnarray}
\Delta s&=&\sqrt{{2\Delta A}\Delta D_{\Omega}s}
=\sqrt{2\Delta A\Delta D_{sC}s}\nonumber\\
&=&\sqrt{2\Delta A\Delta \left(C\,D_ss\right)}
\approx\sqrt{2\Delta A\Delta C\,|D_ss|} \label{uncsexp}\ ,
\end{eqnarray}
provided one assumes $D_ss$ to be approximately constant. The resulting
product of uncertainties $\Delta A$ and $\Delta C$ leads to an
interesting result if one applies a quasi-classical uncertainty
principle to it. That is, to require the product of the uncertainties
of two conjugated observables to be limited by the quantum mechanical
uncertainty principle $\Delta A\Delta C\geq{\hbar}/{2}$, and thus
\begin{equation}
\Delta s \geq \sqrt{\hbar|D_s s|}\label{limitfors}\ .
\end{equation}
A possible interpretation of \eqref{limitfors} could
be that the achievable accuracy within the measurement of
even one single observable is limited,
unlike in established quantum theory where every
  \textit{single} quantity is assumed to be detectable with arbitrary
  accuracy.
This, however, would contradict
the commonly accepted principle of the determinability of a single
observable. A way of overcoming this apparent contradiction was given
by Bohr and Rosenfeld in their work
on the measurability of the quantized electromagnetic field
\cite{BR}. Bohr and Rosenfeld showed that the reason for the apparent
contradiction was that the measurement was not performed
`carefully enough'. They found that the
accuracy improved immediately if a particular term was added to
the total action: when considering a measurement
of the electromagnetic field with the help of a test body, they
found that they had to add another force to the
system. In the language of DeWitt's approach, this means adding a
so-called `compensation term' to the action:
\begin{equation*}
S+\Sigma+g\Omega \rightarrow S+\Sigma+g\Omega-\frac12g^2D_{\Omega}\Omega\ .
\end{equation*}
In virtue of this compensation mechanism, the second term of
\eqref{sexp} drops out and \eqref{uncsexp} becomes
\begin{equation*}
\Delta s=\frac{\Delta A}{g}\ ,
\end{equation*}
which contains no fundamental limitation on the measurement of $s$
anymore.

In fact, for the simultaneous measurement of \textit{two} observables
nothing essentially or conceptually new has to be added. The only
thing that is required is the addition of a further coupling term
corresponding to the measurement of the second quantity $s_2$, say
$\Omega_2$, and the corresponding compensation term to the
action. The total coupling then reads
\begin{equation*}
g(\Omega_1+\Omega_2)-\underbrace{\frac12g^2D_{\Omega_1+\Omega_2}(\Omega_1+\Omega_2)}_{\text{compensation
    term}}\ .
\end{equation*}
To ensure that the resulting equations are resolvable with respect to
experimental data one has, analogously to \eqref{action one obs} and
\eqref{D one obs}, to demand
\begin{eqnarray*}
\Omega_1&=&s_1B_1 \quad\quad\quad\quad\text{with}\quad\quad\quad
(B_1,A_1)=1\ ,\\
\Omega_2&=&s_2B_2 \quad\quad\quad\quad\text{with}\quad\quad\quad
(B_2,A_2)=1\ , \\
D_{\Omega_1}A_1&=&s_1 \ ,\\
D_{\Omega_2}A_2&=&s_2 \ .
\end{eqnarray*}
After the introduction of classical uncertainties for the product of
$\Delta s_1$ and $\Delta s_2$ (minimized with respect to $g$), this
leads to
\begin{equation}
\Delta s_1 \Delta s_2=\frac12\left(\Delta
  A_1\Delta\left(\Omega_1,s_1\right)+\Delta
  A_2\Delta\left(\Omega_2,s_2\right)\right)\label{twoobs}\ .
\end{equation}
Together with
\begin{eqnarray*}
\left(A_1,\left(\Omega_1,s_1\right)\right)&=&-\left(\Omega_1,\left(s_1,A_1\right)\right)-\left(s_1,\left(A_1,\Omega_1\right)\right)=\left(s_1,s_2\right)\
,\\
\left(A_2,\left(\Omega_2,s_2\right)\right)&=&-\left(\Omega_2,\left(s_2,A_2\right)\right)-\left(s_2,\left(A_2,\Omega_2\right)\right)=\left(s_2,s_1\right)\
,\label{relations}
\end{eqnarray*}
it is possible to arrive at a, as DeWitt called it, \emph{universal
  principle}. This is done by assuming the existence of a
\emph{fundamental principle} which limits the accuracy of two,
simultaneously considered, apparatus observables by

\begin{equation}
\Delta A\Delta B\ge\frac{\hbar}{2} |\left(A,B\right)|\label{qmapparatus}\ .
\end{equation}
For example, this could be $A_1$ and $A_2$ and, if independent of the
system trajectories, $\left(\Omega_1,s_1\right)$ and
$\left(\Omega_2,s_2\right)$. This fundamental principle
together with \eqref{twoobs} would result in the corresponding
principle for the system observables
\begin{equation}
\Delta s_1\Delta
s_2\ge\frac{\hbar}{2}|\left(s_1,s_2\right)|\label{result, dewitt2}\ ,
\end{equation}
which could be interpreted as a \emph{universal principle} valid for
any two observables of any system coupled to an apparatus obeying
\eqref{qmapparatus}.

However, one cannot conclude from this inequality that the system that
is being measured \textit{has} to be quantized.
The only place where quantum mechanical considerations appear explicitly
is in the observation that the quantization rule $[A,B]=i\hbar(A,B)$
applied to apparatus observables automatically results in an uncertainty
relation for $A$ and $B$ that satisfies \eqref{qmapparatus}, and therefore
that an apparatus that is quantum mechanical in nature would satisfy his
fundamental principle.
DeWitt's derivation, however, makes use of classical Poisson brackets only,
which indicates that quantum considerations do not play a fundamental
role in his calculation.
His argument is, after all, very general: it applies equally well
to two classical systems that interact, or to a classical system that
interacts with a quantum system, as long as all of his requirements
are satisfied.
Furthermore, one has to recall that the uncertainty relations have
nothing to do with disturbances; they only express the limited
applicability of classical concepts \cite{deco}.
DeWitt's argument therefore does not provide a proof that the quantum
theory must be extended to all physical systems.

\section{Outlook}

Inspired by the gedanken experiment proposed in \cite{EH}, we have
investigated a model in which a gravitational wave interacts
with a mass-cube. In order to discuss the mutual
interaction, we have applied a DeWitt-type measurement analysis to
the coupled system. We have found that if one system possesses
some uncertainty, this uncertainty is necessarily
transferred to the other system. It does not matter
whether the uncertainty is of quantum-mechanical or classical
origin. DeWitt's formalism is general enough to encompass both
situations.
We have also shown that some of the arguments in \cite{EH}
are incorrect because they do not take into account the quantum
entanglement which arises in the situation under study.

We have investigated a second model in which a classical gravitational
wave interacts consistently with a quantized scalar field. This hybrid model
provides a counterexample to the claim that a system coupled to a
quantum system must necessarily also be of quantum nature. We have
considered the argument made in \cite{DeWitt62} that the quantum theory
must be extended to all physical systems, and have shown that this
conclusion is not justified. The universal validity of quantum theory
would in turn necessarily entail the quantization of the gravitational
field, cf. also the remarks by Richard Feynman in this context,
for example: ``\ldots if you believe in quantum
mechanics up to any level then you have to believe in gravitational
quantization \ldots '' \cite{Feynman} . One cannot,
therefore, conclude from these gedanken experiments alone, without
assuming the universal validity of quantum theory, that gravity {\em
  must} be quantized. This is similar to the old analysis by Bohr and
Rosenfeld about the measurability of the quantized electromagnetic
field; this analysis does not show that the electromagnetic field must
be quantized but that the results of the measurement analysis are in
accordance with the quantum electrodynamical commutation
relations, cf. \cite{rosenfeld}. Of course, by empirical arguments
(e.g. the observed coupling of photons to matter) one knows that the
electromagnetic interaction is of quantum nature. As long as
quantum-gravitational experiments are not possible, analogous
empirical arguments are unavailable for gravity; there exists,
however, an experiment that falsifies (under some assumptions) the
standard version of semiclassical gravity where classical gravity is
coupled to the expectation value of a quantum energy-momentum tensor
for matter \cite{PG}.

 From
general arguments (singularity theorems, universality of gravity,
unification), the quantization of gravity seems unavoidable
\cite{OUP}, although there is no logical proof.
The task is then to consider particular approaches to quantum gravity,
such as loop quantum gravity or string theory, and to perform a
quantum measurement analysis along the lines of \cite{BR} or
\cite{DeWitt62}. This is, however, left to future publications.

\section*{Acknowledgements}

M.R. would like to thank Michael J. W. Hall for very helpful
discussions. M.A. gratefully acknowledges stimulating discussions with
Friedemann Quei\ss er and Barbara Sandh\"ofer. C.K. and M.R. are,
moreover, most grateful to Steve Carlip for a helpful and
illuminating correspondence.

\appendix
\section{Application of the mixed classical-quantum system to a simple
oscillator model}

We consider here the interpretation of the solution $\{ S^c\left[
  \phi,t \right), P^{c }\left[ \phi,t \right)\}$ for the case $g
\rightarrow 0$ which we derived in Sec. \ref{cl-quant.systems}.

It will be helpful to consider first a simpler but closely related
system, an ensemble in configuration space for the classical
one-dimensional harmonic oscillator. The equations that describe the
state of the system are the Hamilton--Jacobi equation and the
continuity equation,
\begin{eqnarray}
\dot{S}_{osc}+\frac{1}{2}\left( \frac{\partial S_{osc}}{\partial x}\right)
^{2}+\frac{1}{2}\omega ^{2}x^{2} &=&0, \nonumber\\
\dot{P}_{osc}+\frac{\partial }{\partial x}\left( P\frac{\partial S_{osc}}{%
\partial x}\right)  &=&0.  \nonumber
\end{eqnarray}%
The Hamilton--Jacobi equation has a solution of the form%
\begin{equation}
S_{osc}\left( x,t\right) =-\frac{\omega x^{2}}{2}\tan \left(\omega
  t\right). \nonumber
\end{equation}%
Given $S_{osc}$, the most general solution of the continuity equation
is given by
\begin{equation}
P=\frac{1}{\cos \left( \omega t\right) }f\left( \frac{x}{\cos \left(
      \omega t\right) }\right), \nonumber
\end{equation}%
where $f$ is an arbitrary function. In particular, there is a Gaussian
solution,
\begin{equation}
P_{osc}\left( x,t\right)= \sqrt {\frac{\tau}{2\pi \cos^2 \left( \omega
      t\right) }} \exp \left\{ -\frac {1} {2} \frac{\tau}{\cos^2
    \left( \omega t\right) }\left( x- w \cos \left( \omega t\right)
  \right) ^{2}\right\}, \nonumber
\end{equation}%
where $w$ and $\tau$ are constants.

At $t=0$, the state has average momentum
\begin{equation}
\langle p \rangle (0) \equiv \int \; dx \; P(x,0) \frac{\partial S_{osc}}{\partial x}(x,0) = 0 \nonumber
\end{equation}%
and no momentum uncertainty (i.e., $\Delta p(0)=0$); average position
$\langle x \rangle (0) = w$ and position uncertainty $\Delta x(0) =
\tau^{-1/2}$. The solution therefore represents a state which is
prepared with zero momentum and no momentum uncertainty, but localized
in space. Furthermore, when
$\omega t \rightarrow \{ \pm \frac{\pi}{2},\pm \frac{3\pi}{2},...\}$,
$P_{osc}\left( x,t\right)$ becomes a delta function and the particle is
found at $x=0$ with probability one. The total energy of the state is
given by
\begin{equation}
E=\langle-\frac{\partial S_{osc}}{\partial t}\rangle  =
\frac{1}{2}\omega ^{2}\tau^{-1}. \nonumber
\end{equation}%

A formal solution of the \emph{quantum} analogue of this problem is
obtained by taking the limit $\tau \rightarrow 0$. One can check that
the wave function $\psi = \sqrt{P_{osc}} \;e^{iS_{osc}/\hbar}$ is then
a solution of the Schr\"{o}dinger equation, and that it corresponds to
a state which is prepared with zero momentum and no momentum
uncertainty but completely delocalized.

Consider now the solution $S^c\left[ \phi,t \right)$ and $P^{c }\left[
  \phi,t \right)$ of Sec. \ref{cl-quant.systems}. To see the
analogy to the one-dimensional harmonic oscillator, evaluate
$S^c\left[ \phi,t \right)$ and $\frac{\delta S^c\left[ \phi,t
  \right)}{\delta \phi _{x}}$ using the representation $\phi
_{x}=\sum_{k}a_{k}f_{x}^{(k)}$and $F_{yx}=-\sum_{k} k\tan \left(
  kt\right)f_{y}^{(k)}f_{z}^{(k)}$, where the $f_{y}^{(k)}$ are the
basis functions introduced previously. This leads to
\begin{eqnarray}
S^c\left[ \phi,t \right) &=& -\sum_{k}\frac{k a_{k}^{2}}{2}\tan \left(
  kt\right)  \nonumber\\
\frac{\delta S^c\left[ \phi,t \right)}{\delta \phi _{x}} &=& -\sum_{k}
k a_{k} \tan \left( kt \right)f_{x}^{(k)} \nonumber.
\end{eqnarray}%
Furthermore, $P^{c}\left[ \phi,t \right) $ is given by
\begin{equation}
P^{c}\left[ \phi,t \right)= N^{c}(t)\;e^{ -\frac{1}{2}\int \int
dadb\;\left( \phi _{a}-\beta _{a}\right) K_{ab}(t)\left( \phi _{b}-\beta
_{a}\right) } \nonumber
\end{equation}%
with%
\begin{eqnarray}
K_{yx} &=& \sum_{k}\frac{\tau_k}{ \cos ^{2} \left( kt\right) }f_{x}^{(k)}f_{y}^{(k)}, \nonumber\\
\beta _{x}&=&\sum_{k} w_k \cos(kt) f_{x}^{(k)}, \nonumber\\
N^{c } &\sim&  \frac {1} { {\Pi_{k}\cos \left( kt\right) }}. \nonumber
\end{eqnarray}
Comparing these expressions, one can see that the solution $S^c \left[
  \phi,t \right) $ is analogous to $S_{osc}\left( x,t\right)$ while
$P^{c}\left[ \phi,t \right)$ is analogous to $P_{osc}\left(
  x,t\right)$.

Just as in the case of the one-dimensional harmonic oscillator, a
formal solution of the corresponding quantized field equations is
obtained by taking the limit $\tau \rightarrow 0$.

\section{ Scattering of classical and quantum non-relativistic
  particles that interact gravitationally}

Consider a gedanken experiment that involves the scattering of two
non-relativistic particles, one a classical particle of mass $M$ (the
projectile) and the other one a quantum particle of mass $m$ (the
target). The interaction is assumed to be caused by the gravitational
attraction between the two particles.

This gedanken experiment is most interesting when the initial
amplitude for the quantum particle (i.e., as $t \rightarrow -\infty$
when the two particles are very far from each other so that the
interaction term can be neglected) has two peaks of equal magnitude, A
and B, that are well separated. Then, in a measurement,
the quantum particle will be
``found'' at the location of peak A (with probability 1/2) or at the
location of peak B (with probability 1/2). (In the Everett
interpretation, the measuring agency is entangled with this separate
possibilities.)
Consider the case where the
classical particle comes very close to peak A and remains at all times
at a very large distance from peak B.

What happens when the classical particle scatters from the quantum
particle? A ``naive approach'' suggests three possible mutually
exclusive outcomes: (a) the quantum particle is found at the location of
peak A and the classical particle comes very close to the quantum
particle of mass $m$: the scattering is very strong; (b) the quantum
particle is found at the location of peak B and the classical particle never
comes very close to the quantum particle of mass $m$: the scattering
is very weak; (c) as in ``semiclassical gravity'',
cf. \cite{PG}, the classical
particle ``sees'' a mass $m/2$ at the location of peak A and it
``sees'' a mass $m/2$ at the location of peak B: the scattering is
about one half of what one would calculate under assumption (a).

Each of these possibilities seems unrealistic. The source of
difficulties is, clearly, the use of the ``naive approach'': it is
impossible to derive any reasonable conclusion without introducing a
concrete model for this mixed classical-quantum system. We show below
that none of these outcomes is supported by a more careful analysis.

The formalism of configuration space ensembles allows a general and
consistent description of interacting classical-quantum systems
\cite{HR}, \cite{harxiv}, and we will now apply it to this problem.
Using this formalism, it is straightforward to set up the equations
that are needed to describe the gedanken experiment. Let $q$ denote
the configuration space coordinate of a quantum particle of mass $m$,
and $x$ denote the configuration coordinate of a classical particle of
mass $M$, and consider an interaction term of the form
\begin{equation} \label{cqvg}
V(q,x) = -G\frac{mM}{|x-q|}.\nonumber
\end{equation}
Within this formalism, the equations that describe the system are
derived from the Hamiltonian
\begin{equation} \label{cqhamil}
\tilde{H}_{QC}[P,S] = \int dq\,dx\, P\,\left[ \frac{|\nabla_q S|^2}{2m}
+ \frac{|\nabla_x S|^2}{2M} -G\frac{mM}{|x-q|} + \frac{\hbar^2}{4} \frac{|\nabla_q \log P|^2}{2m} \right]
\end{equation}
and take the form
\begin{equation} \label{cqcontg}
\frac{\partial P}{\partial t} =  -\nabla_q .\left( P \frac{\nabla_qS}{m} \right)
- \nabla_x.\left(P\frac{\nabla_x S}{M}\right) ,
\end{equation}
\begin{equation} \label{cqhjg}
-\frac{\partial S}{\partial t} =
\frac{|\nabla_qS|^2}{2m} + \frac{|\nabla_xS|^2}{2M}
- \frac{\hbar^2}{2m}\frac{\nabla_q^2 P^{1/2}}{P^{1/2}} - G\frac{mM}{|x-q|}.
\end{equation}
Equations of this type have been studied in detail in references
\cite{HR} and \cite{harxiv} and we refer the reader to these papers;
here we summarize a few aspects of the formalism. Moreover, Sect.~IV
presents a detailed discussion of a model based on this approach.

The state of the mixed classical-quantum system is
described by two functions, $P(x,q,t)$ and $S(x,q,t)$, which have the
following physical interpretation: $P$ is a probability density
defined over configuration space, and $\frac{1}{M} P \nabla_x S$
($\frac{1}{m} P \nabla_q S$) is a probability current associated with the
classical (quantum) particle. Equation (\ref{cqcontg}) has the form of a 
continuity equation and (\ref{cqhjg}) has the form of a modified
Hamilton-Jacobi equation. One may introduce a wave function of the
form $\psi=\sqrt{P}e^{iS/\hbar}$ which will satisfy a non-linear
generalization of the Schr\"{o}dinger equation.

Already the equations show some features that tell us what to expect
of the solutions:

(1) When
$|\frac{\hbar^2}{2m}\frac{\nabla_q^2 P^{1/2}}{P^{1/2}}| << |G\frac{mM}{|x-q|}|$,
we can neglect the term proportional to
$\frac{\hbar^2}{2m}\frac{\nabla_q^2 P^{1/2}}{P^{1/2}}$ in (\ref{cqhjg}).
This will be the case when the mass $m$ of the quantum system is large
enough for this inequality to be valid. But if this is the case,
(\ref{cqhjg}) reduces to the classical Hamilton-Jacobi equation and
we end up with the equations of a classical-classical system. This
shows that the formalism has the correct classical limit.

(2) If the interaction term $V(q,x) = -G\frac{mM}{|x-q|}$ that appears
in (\ref{cqhjg}) is very small (say at $t \rightarrow -\infty$ when
the two particles are very far
from each other) and there is no initial correlation between
the particles, then the non-linearity in the equations of
the quantum particle will amount to only a small perturbation and the
superposition principle will be valid for the quantum sector to a very
good approximation. This means that the formalism has the correct
quantum limit. Notice, however, that the equations are non-linear when
the interaction term is taken into consideration: the quantum
superposition principle breaks down when the interaction between the
classical and quantum particles is strong.

(3) Suppose that the systems were independent before the interaction
(i.e., at $t \rightarrow -\infty$ when the two particles are very far
from each other). That amounts to postulating initial conditions
\begin{eqnarray}
P^{\;(-\infty)}(x,q,t) &=& P^{\;(-\infty)}_C(x,t) P^{\;(-\infty)}_Q(q,t)\nonumber\\
S^{\;(-\infty)}(x,q,t) &=& S^{\;(-\infty)}_C(x,t) \; + \; S^{\;(-\infty)}_Q(q,t)\nonumber
\end{eqnarray}
Before the interaction, the combined classical-quantum system breaks
up naturally into classical and quantum sectors. However, after the
interaction, the two systems will not be independent anymore, and we
will have
\begin{eqnarray}
P^{\;(+\infty)}(x,q,t) &\neq& P^{\;(+\infty)}_C(x,t) P^{\;(+\infty)}_Q(q,t)\nonumber\\
S^{\;(+\infty)}(x,q,t) &\neq& S^{\;(+\infty)}_C(x,t) \; + \; S^{\;(+\infty)}_Q(q,t)\nonumber
\end{eqnarray}
Now the combined classical-quantum system will no longer have well
defined classical and quantum sectors. This means that a measurement
of the position of either the classical or quantum particle will force
a change of the fields $\{P(x,q,t),S(x,q,t)\}$ that describe the
\emph{total} system. In other words, the classical and quantum sectors
have become entangled.

(4) Cases (a) and (b) of the ``naive approach'' are ``either-or''
outcomes that implicitly assume that there is no entanglement in the
case of mixed classical-quantum systems. But we have just shown that a
consistent theory of interacting classical-quantum systems leads to
final states that \emph{are} entangled. We have to conclude therefore
that these two outcomes are not supported by a more careful analysis
and that they must be rejected. Furthermore, the theory that we have
used is fundamentally different from standard ``semiclassical
gravity''. Therefore, case (c) of the ``naive approach'' is also
excluded.

The predictions of the mixed classical-quantum system described by
({\ref{cqcontg}) and ({\ref{cqhjg}) differ then substantially from
the outcomes predicted using the ``naive approach''. The qualitative
features of the solution can be determined without carrying out a
detailed calculation. To do this, we introduce centre-of-mass
and relative coordinates
\[  \overline{x}:=\frac{mq+Mx}{m+M},~~~~~r:= q-x .\]
and the total mass $M_T$ and relative mass $\mu$ defined by
\begin{equation}
M_T := m+M,~~~~~~\mu:= \frac{mM}{m+M}.\nonumber
\end{equation}
Rewriting the Hamiltonian ({\ref{cqhamil}) in terms of these new
coordinates leads to \cite{HR}
\begin{eqnarray}\label{cqhamilcm}
\tilde{H}_{QC} &=& \int d\overline{x}dr \, P \left[ \frac{
|\nabla_{\overline{x}} S|^2}{2M_T} + \frac{\hbar^2m}{4(m+M)} \frac{|\nabla_{\overline{x}} \log P|^2}{2M_T} \right]\nonumber\\
~~&~&~~ +\int d\overline{x}dr \, P \left[ \frac{
|\nabla_r S|^2}{2\mu} + \frac{\hbar^2M}{4(m+M)} \frac{|\nabla_r \log P|^2}{2\mu}
-G\frac{\mu M_T}{|r|} \right]\nonumber\\ 
~~&~&~~ - \frac{\hbar^2}{4(m+M)} \int d\overline{x}dr \, \frac{ \nabla_{\overline{x}}P.\nabla_r P}{P}  .
\end{eqnarray}
To interpret this expression, compare to the Hamiltonian of a purely
quantum system, which is of the general form \cite{HR}
\begin{equation}
\tilde{H}_Q[P,S] =   \int dq\, P \left[ \frac{|\nabla S|^2}{2m}
+ \frac{\hbar^2}{4} \frac{|\nabla \log P|^2}{2m} + V(q) \right] \nonumber.
\end{equation}
We see then that the Hamiltonian $\tilde{H}_{QC}$ in ({\ref{cqhamilcm}) 
is the sum of three terms: (i) a quantum-like term corresponding 
to free centre-of-mass motion but with a rescaled Planck constant
\[ \hbar_{\overline{X}} := [m/(m+M)]^{1/2}\, \hbar ; \]
(ii) a quantum-like term corresponding to relative motion in
a potential $V(r)=-G\frac{\mu M_T}{|r|}$ but with a rescaled Planck constant
\[ \hbar_R := [M/(m+M)]^{1/2}\, \hbar ; \]
and (iii) an intrinsic interaction term. One would expect then a solution
with \emph{qualitative} features that resemble those of a purely
quantum system, however with important modifications induced by the
(iii) term and the rescaling of the Planck constant in (i) and (ii).
Regarding conservation laws, we point out that the equations of motion
(\ref{cqcontg}) and (\ref{cqhjg}) are invariant under Galilean transformations
 \cite{HR}, therefore the usual conservation laws that follow from Galilean
 invariance apply. A more detailed analysis of this particular mixed
 classical-quantum system is in preparation.

(5) As mentioned before, the wavefunction $\psi$ associated with this
mixed classical-quantum system obeys a non-linear generalization of
the Schr\"{o}dinger equation and the theory can be seen as a
non-linear modification of quantum mechanics. Therefore, the concepts
that are used to describe measurements in quantum theory will have
their counterparts in this formalism, although perhaps with limited
validity. For example, if we want to modify standard quantum theory by
introducing a wavefunction collapse
(i.e., a discontinuous change in the
wavefunction due to an observation), then we will need to assume that
there is something equivalent here that applies to the \emph{whole}
system (i.e., a discontinuous change in $\{P(x,q),S(x,q)\}$ due to an
observation).



\end{document}